\documentclass[journal=jpccck,manuscript=article]{achemso}
\usepackage{times}
\usepackage{epsfig}
\usepackage{graphicx}
\usepackage{psfrag}
\usepackage{ae}
\usepackage{amsmath,amssymb}
\usepackage[usenames]{color}
\usepackage{float}

\title{Sodium Chloride, NaCl/$\epsilon$: New Force Field}

\author{ Ra\'ul Fuentes-Azcatl} 
\email{razcatl@hotmail.com}
\affiliation{Instituto 
de F\'{\i}sica, Universidade Federal
do Rio Grande do Sul, Caixa Postal 15051, CEP 91501-970, 
Porto Alegre, RS, Brazil}
\author{Marcia C. Barbosa} 
\email{marcia.barbosa@ufrgs.br}
\affiliation{Instituto 
de F\'{\i}sica, Universidade Federal
do Rio Grande do Sul, Caixa Postal 15051, CEP 91501-970, 
Porto Alegre, RS, Brazil}

\begin{document}

\begin{abstract}
A new computational 
model for sodium chloride, the NaCl/$\epsilon$, is proposed.
The force field employed for the description of the NaCl
 is based on a set of radial particle-particle pair 
potentials involving Lennard-Jones (LJ) and Coulombic forces.
The parametrization is obtained by fitting 
the density of the crystal and 
the density and the dielectric constant 
of the mixture of the salt with water at a diluted solution.
Our model shows good agreement with the experimental values 
for the density and for the surface tension 
of the pure system and for the 
density, the  viscosity, the diffusion, 
and the dielectric constant for the mixture
with water at  various molal
concentrations. 
The NaCl/$\epsilon$ together with 
the water TIP4P/$\epsilon$ models
provide a good approximation
for studying electrolyte solutions.
\end{abstract}

\maketitle

\section{Introduction}

Sodium chloride is present in our lives 
from the chemical balance of our body to 
the geophysical and biological equilibrium 
of the planet. It
 is also largely used in industry, particularly to preserve food. 
Therefore,  the understanding of 
physical-chemical properties of  sodium chloride
as a pure substance or in mixtures is 
important.One of the key questions
regarding salt is how 
it behaves in solution under different
pressures and temperatures and also 
under confinement~\cite{ Manning, Auffinger, Klein, Hovland}. 

A number of experimental studies have
addressed the behavior of  sodium chloride in water~\cite{CRC,Kuma}.
Even though they provide the behavior
of the thermodynamic and of the
dynamic quantities as a function of temperature and pressure,
due to the high number of variables that influence 
these properties, it becomes difficult 
to identify which is the  mechanism behind the behavior
of the salt solutions.
Then, the  theoretical methods become  a complementary
tool that not only allows for exploring a 
wider range of parameters but also provide a 
more controled analysis of the  
parameters.
 Due to the 
long range nature of
the Coulomb interactions, analytic approaches
for describing the behavior of the ions, $Na^+$ and $Cl^-$,
in water
require approximations 
that either limit the analysis to very low dilution~\cite{Debye-Huckel} or
to the study of  systems far from phase separations~\cite{MSA}.
 Consequently   after the development
of approaches
to account for the electrostatic
interactions~\cite{evaltsum,allen}
simulations became an important
strategy to study electrolyte solutions.

The crucial step in the simulations
is to construct
an appropriated force field for the interaction
potential between the ions 
and the water. The usual method is to
fit the parameters
of the  model  with the experimental 
results for  the density
and for the structure for the
real system  at one
 determined  pressure
and temperature. Then, the results
obtained for thermodynamic and dynamic properties with the
 model are compared with experiments. 
Following this procedure, 
a number of models for sodium chloride~\cite{Moucka13NaCl} 
capable of reproducing the density of the pure system have been developed.

Recently Smith W.R. et al~\cite{Moucka13NaCl} studied 
thirteen of the most common  NaCl force fields. These
models, even thought reproduce some of the properties
of the crystal, are unable to capture others. For instance, just one of them 
reproduced the correct density  and  other  obtained  the 
correct  chemical 
potential of the solid phase at room temperature. 
In parallel to 
modeling the salt,  numerical strategies have  been employed to 
build  computational models for water~\cite{SPC,SPCE,TIP4PQ,TIP4P05}.
These models reproduce the density of water 
around $298\;K$ 
and a pressure of $1$bar~\cite{vega11} but fail to provide a
reliable value for a number of properties, including
the dielectric constant~\cite{TIP4Pe}.

In the case of studying electrolyte solutions, 
the common strategy is to combine one 
model for water and one model for salt that 
have been obtained by fitting the properties
of the pure systems.
Then, the mixture of these two models is tested.
The solubility is one of the main
properties used to validate the model of salt.
 When dissolved in water, the molecule of 
 sodium chloride  dissociates in one cation,  $Na^+$, and 
one anion,$ Cl^-$. Due to the polar character of the water molecules 
both ions become  
surrounded by water molecules~\cite{CRC}. For certain
salt concentrations the
system phase separates in a salt rich phase and 
salt poor phase. The solubility can be
computed in this coexistence.

 One method for
computing the coexistence
between the crystal  and
the saturated phase
is to estimate the chemical
potentials independently~\cite{Ferrario, Sanz}. For 
the solid 
 the absolute free
energy of the crystal can be
computed  using the method proposed by Frenkel and
Ladd ~\cite{Frenkel}.
Employing this framework, Sanz and Vega~\cite{Sanz} 
 determined the solubility of KF and NaCl in the water solution. 
This procedure was also employed
for a variety of salts in water~\cite{Lisal,Mouka11,Mouka12}
not only for computing the solubility but also other
properties of the salt solution~\cite{Pa14,Ga10}.
Within the same method the best comparison
 between the 
experimental values for the solubility
and the simulations was computed by  
Smith and coworkers~\cite{Moucka13NaCl}.

Another approach to obtain the salt rich and salt poor
coexistence
is to use a sufficiently large crystal
in contact with an almost saturated ion solution~\cite{Jo09}. The
main assumption is that this crystal
and the solution reach an
equilibrium state after very long
simulations and extremely large systems, otherwise
finite size effects dominates~\cite{Hector}.

The drawback of analyzing the water and 
salt mixture using the NaCl and water
models parametrized for the pure systems 
is that when mixed, the water surrounds 
the ions what affects the salt-salt and water-water 
interactions.
In this work we present a new model for NaCl that 
is parametrized to 
reproduce  properties of the pure salt and of
the water-salt mixture. The
 behavior of the model is tested 
against experiments for properties
of the pure salt and  of the 
water-salt solution. In the
salt-water mixture, two water models designed 
to give the correct dielectric constant
of water were 
employed:  the SPC/$\epsilon$ \cite{spce} and 
the TIP4P/$\epsilon$\cite{TIP4Pe}. 
For computing the solubility we follow the 
Manzanilla  et. al~\cite{Hector} approach
with the crystal surrounded by the saturated solution
to avoid finite size effects.

The remaining of the 
paper goes as follows. In the section 2 
the new model for NaCl is introduced  and the two water models
employed for the parametrization of the salt
were  reviewed. The section 3 summarizes
the simulation details
and  the results are analyzed in the Section 4. The conclusions
are presented in the section 5.

\section{The Models}

\subsection{The NaCl/$\epsilon$ Model}

The force field employed here for the description of the NaCl  
in the aqueous solution is based on a set of radial particle-particle pair 
potentials involving Lennard-Jones (LJ) and Coulombic contributions, namely
\begin{equation}
\label{ff}
u(r_{ij}) = 4\xi_{ij} 
\left[\left(\frac {\sigma_{ij}}{r_{ij}}\right)^{12}
-\left (\frac{\sigma_{ij}}{r_{ij}}\right)^6\right] 
+ \lambda_i\lambda_j\frac{q_iq_j}{4\pi\epsilon_0r_{ij}}
\end{equation}
\noindent where $r_{ij}$ is the distance between sites $i$ and $j$, $q_i$ is 
the electric charge of site $i$, $\epsilon_0$ is the permittivity of 
vacuum,  $\xi_{ij}$ is the the potential depth and  $\sigma_{ij}$ the 
distance at which the interparticle potential is zero.

We assume that the pure water and the ion 
potentials are compatible. This means that
  the cross interactions between 
the water molecules and the ions can be calculated by the Lorentz-Berteloth 
(LB) combing rules for the conformal LJ potential, \cite{Hansen}
\begin{equation}
\label{LB}
\sigma_{\alpha\beta}=\bigg(\frac{\sigma_{\alpha\alpha}
+\sigma_{\beta\beta}}{2}\bigg)\ \ ;\ \ \xi_{\alpha\beta}
=\sqrt {\xi_{\alpha\alpha}\xi_{\beta\beta}}\;.
\end{equation}

For the NaCl/$\epsilon$ model the NaCl is considered an ion pair,
 $\xi_{ij}=\xi_{LJ}$
while  $\sigma_{ij}=\sigma_{LJ}$
for any $i$ and $j$ namely  for
Na-Na, Cl-Cl or Na-Cl~\cite{SD94}.
 The spherical anions and cations are represented
by a single interactive site at 
their centers, carrying charges $q_{i}=\pm 1\;e$
where $e$ is the charge of an electron.
In order to correct for the nonpolarizability of 
the model the Coulombic term is corrected
by a screening factor $\lambda_i=\lambda_C$ for 
both sodium and chloride ions. 
This
factor is used both in the pure salt system
and in the solution with water. 
Therefore, there are
 three parameters, namely $\lambda_{C}$, $\sigma_{LJ}$ and  $\xi_{LJ}$
to be adjusted with experimental data for each ion. 
The assumption that polarizable models for some 
temperature and pressures can be reduced to
simple nonpolarizable models was introduced by
Leontyev and Stuchebrukhov~\cite{Le14} for the particular
case of ionic liquids. Here we explore this
idea for the screening of the salt ions.

The parametrization process was made as follows.
First, the parameters were selected so that the NaCl/$\epsilon$ 
force field  reproduces  the 
 experimental value for the  density 
of the crystal in the face centred cubic phase
at $1$bar and $298\;K$~\cite{CRC,JC08}.
There are several parametrization
of the NaCl model that give the proper density value. 
A table with all these values,
including the parameters used by other models,
was made. Next, all the possibilities were checked 
with the  radial distribution function, g(r), and 
a subset of parameters which give the correct
density and also describe
the structure of the 
salt crystal at $1$bar and $298\;K$ were selected. This step provided the 
first approximation for the parameters of the model.

Then, the parameters which give the density
and the structure were tested for  the 
 density and the dielectric constant in the mixture of
the salt with water~\cite{CRC} at $1$bar, $298\;K$  
and $4$ molal salt concentration. At this 
concentration    the ions are hydrated and there is not clusters
 starting a 
nucleation. 
Finally, the  parameters for the  NaCl/$\epsilon$ model 
found through this process are
 shown in the Table~\ref{NaClParam}.

 \begin{table}[h]
\caption{Force field parameter of NaCl/$\epsilon$. }
\label{NaClParam}
\begin{center}
\begin{tabular}{|ccccc|}
\hline\hline
Model  & q/e&$\lambda_{C}$& 
$\sigma$/\AA & $(\xi/k_B)$/K\\
\hline
Na&+1&0.885&2.52&17.44\\
Cl&-1&0.885&3.85&192.45\\
\hline
\end{tabular}
\end{center}
\end{table}
 
\subsection{TIP4P/$\epsilon$ Water Model}

\begin{figure}
\psfig{figure=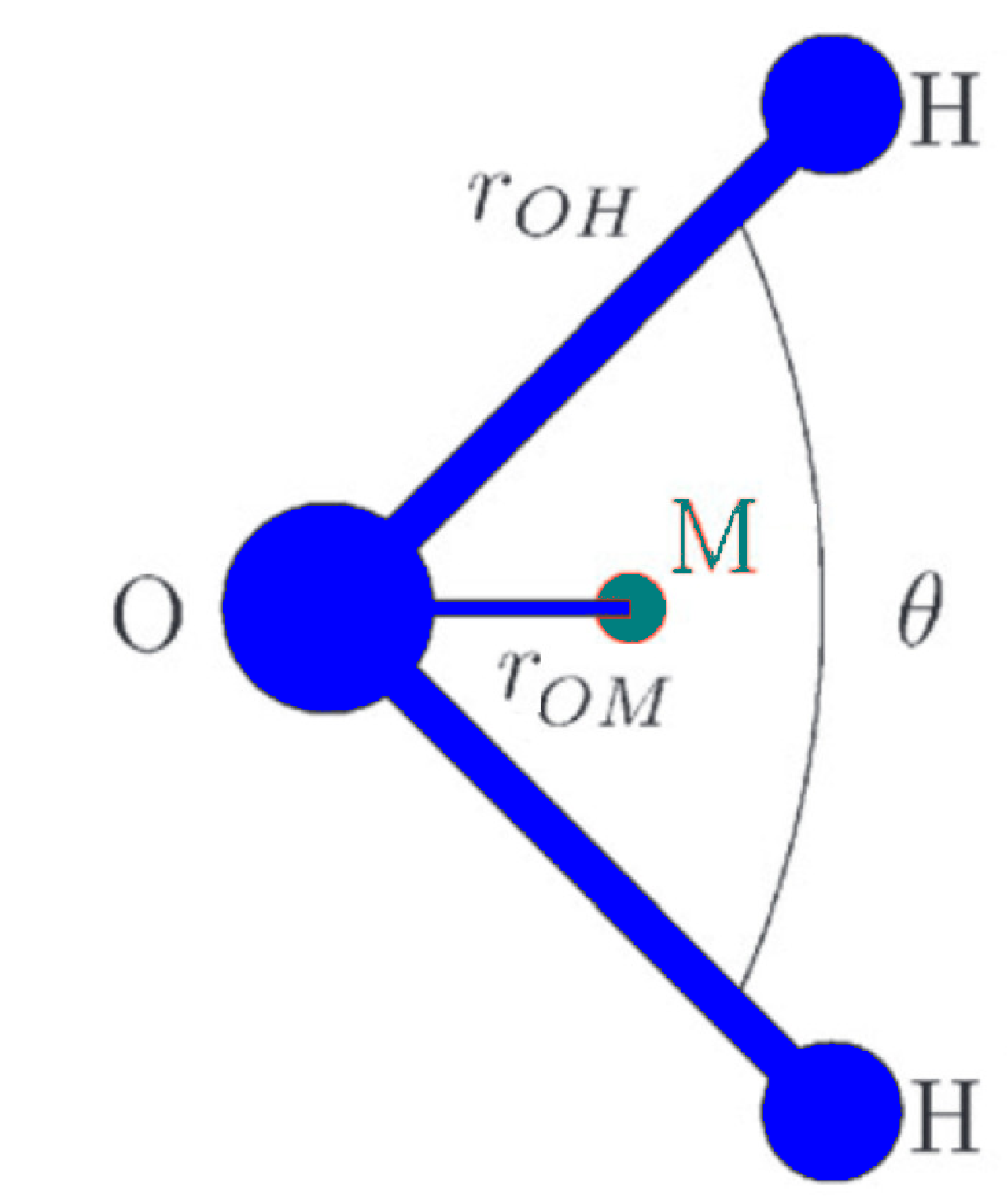,clip,width=4.0cm,angle=0}
\caption{Schematic representation of the TIP4P water model.
The distance between the oxygen and 
the hydrogen is $r_{OH}$ and 
the angle between the oxygen
and the two hydrogens is $\theta$. The hydrogens
have positive charge while the negative charge
is located at a point M $r_{OM}$ distant
from the oxygen that contains no charge. }
\label{fig:TIP4Pe}
\end{figure}

The TIP4P/$\epsilon$~\cite{TIP4Pe}  model defines 
the water molecule as rigid, non-polarizable 
with the same geometry of the TIP4P~\cite{TIP4P} as illustrated
in the Figure~\ref{fig:TIP4Pe}. The intermolecular force 
field between two water molecules is given by the Lennard-Jones and the 
Coulomb 
interactions as given
by the Eq.~\ref{ff}. 
The TIP4P models have a positive
charge at each hydrogen and 
 a negative charge along the bisector of the HOH 
angle located at a distance $l_{OM}$ of the 
oxygen as shown in the Figure~\ref{fig:TIP4Pe}. The geometry and parameters 
of the force fields for the TIP4P/$\epsilon$ are  given in the 
Figure~\ref{TIP4Pe}. In the case of the  TIP4P/$\epsilon$ model
the $\lambda_O=\lambda_H=1$ in the Eq.~\ref{ff} \\

\begin{table}[h]
\caption{Force field parameters of TIP4P/$\epsilon$ water model. The 
charge in site $M$ is $q_M=-(2 q_H)$.  }
\label{TIP4Pe}Table~
\begin{center}
\begin{tabular}[h]{|cccccccc|}
\hline
\hline Model & $r_{OH}$/\AA  & $\Theta$/ $^0$  & $q_H/e$  & $q_M/e$& 
$r_{OM}$ /\AA  & 
$\sigma$/\AA & $(\xi/k_B)$/K\\
\hline

TIP4P/$\epsilon$ & 0.9572 & 104.52 &0.527& 1.054& 0.105 &  3.165 
& 93\\

\hline
\end{tabular}
\end {center}
\end{table}

\subsection{SPC/$\epsilon$ Model}

\begin{figure}
\psfig{figure=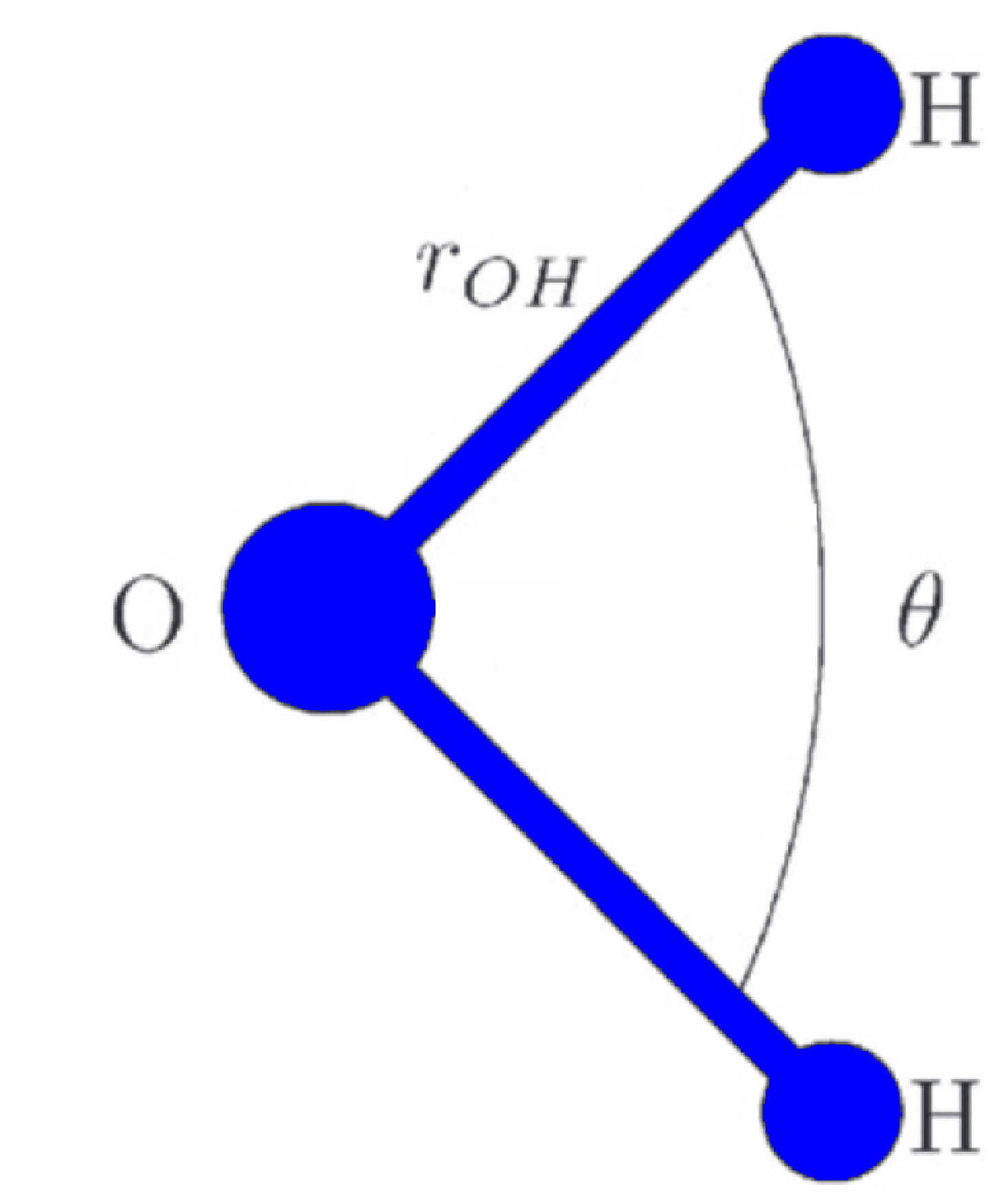,clip,width=4.0cm,angle=0}
\caption{Schematic representation of the SPC water model.
The distance between the oxygen and 
the hydrogen is $r_{OH}$ and 
the angle between the oxygen
and the two hydrogens is $\theta$. The hydrogens
have positive charge while the oxygen carries 
the negative charge. }
\label{fig:spce}
\end{figure}

The SPC/$\epsilon$ is another model for water. It 
is based on the  SPC model geometry shown
in the  Figure~\ref{fig:spce},  but 
with a different set of parameters.
The SPC/$\epsilon$ model~\cite{spce} defines
water as a rigid and  non-polarizable  as illustrated
in the Figure~\ref{fig:spce}. The intermolecular force field between two 
water molecules is given by the  Lennard-Jones and the Coulomb interactions
as given by Eq.~\ref{ff} with $\lambda_O=\lambda_H=1$. 
The parametrization was made using the dipole moment of the minimum 
density method $\mu_{md}$~\cite{TIP4Pe}.

The  SPC/$\epsilon$ model
gives similar thermodynamic and dynamic properties as the  
SPC~\cite{SPC} and the SPC/E~\cite{SPCE}models, but a
better agreement with the experiments for 
the  dielectric constant~\cite{fernandez}.

The geometry and parameters of the force fields
 analyzed in this work are given in the Figure~\ref{spcepsilon}.\\

\begin{table}[h]
\caption{Force field parameters of water model, SPC/$\epsilon$.The 
charge of Oxygen is $q_O=-(2 q_H)$.}
\label{spcepsilon}
\begin{center}
\begin{tabular}{|cccccc|}
\hline\hline
 Model & $r_{OH}$/\AA  & $\Theta$/ $^0$ & $q_H/e$  &   $\sigma$/\AA & 
$(\xi/k_B)$/K\\
\hline
SPC/$\epsilon$ & 1 & 109.45 &0.445&    3.1785 & 84.9\\
\hline
\end{tabular}
\end{center}
\end{table}

\section{The Simulation Details}

Molecular dynamic (MD) simulations were performed using 
GROMACS~\cite{gromacs}(version 4.5.5.). The equations of motion were
solved using the leap-frog algorithm~\cite{allen,gromacs} and 
the time step was $2\; fs$, the time of simulations of different molalities
 is 30 ns, 
keeping the positions and velocities for every 500 steps in simulation. The 
calculus 
of the shear viscosity, however, employed $1\;fs$,the time of simulations is 
40 ns 
and storing the positions and velocities every simulation step. Ewald 
summations were used to
 deal with electrostatic contributions. The real part of the Coulombic
 potential is truncated at 10\AA. The Fourier component of the Ewald 
sums was evaluated by using the smooth particle mesh Ewald (SPME) 
method~\cite{SPME} using a grid spacing of 1.2\AA and a fourth degree
 polynomial for the interpolation. The simulation box is cubic 
throughout the whole simulation and the geometry of the water molecules
 kept constant using the LINCS procedure \cite{LINCS}. Temperature has 
been set to the desired value with a Nos\'e Hoover thermostat~\cite{nhc}. The 
pressure is obtained using the Parinell-Rahman barostat with 
a $\tau_P$ parameter of 1.0 ps\cite{gromacs}.

The MD simulations of pure NaCl made in the  NPT ensemble  were carried out 
under $1\;bar$ pressure condition, on a system of $1024$ NaCl pairs, with a 
time step $\bigtriangleup t = 2\;fs$, the time of simulations is 10 ns 
and storing the positions and velocities every 1000 simulation step. 
The coexisting liquid and vapor phases of NaCl were analyzed
in the NVT ensemble on a system of $2916$ NaCl pairs in an elongated 
simulation cell of
dimensions Lx = Ly = 3*Lz, the time of simulations is 10 ns 
and storing the positions and velocities every 500 simulation step and using a 
r$_{cut}$ =2.6nm. The densities of the two phases were
extracted from the statistical averages of the liquid and vapor 
limits of the density profiles~\cite{alj95}. The corresponding surface 
tension $\gamma$ of one planar interface was calculated 
from the mechanical definition of $\gamma$ ~\cite{alj09}
\begin{equation}
\label{Ec3}
\gamma  = 0.5L_{z}[͓͗P_{zz} − 0.5P_{xx} + P_{yy}]
\end{equation} 
where $P_{\alpha\alpha}$ are the diagonal elements of the microscopic
pressure tensor.The factor 0.5 outside the squared brackets takes into 
account the two symmetrical interfaces in the system.\\

For sodium chloride, NaCl/$\epsilon$ in water, the simulations 
have been done using 864 molecules in the isothermal-isobaric
 ensemble NPT, in 
liquid phase at different molalities and a temperature of $298\;K$ 
and $1\;bar$ of 
pressure. The molality concentration is obtained 
from the total number of ions in solution $N_{ions}$ , the number of 
water molecules $N_{H_{2}O}$ and the molar mass of water $M_{H_{2}O}$ as:\\
 \begin{equation}
\label{mol}
\left[NaCl \right] =\frac{N_{ions}\times 10^{3}}{2N_{H_{2}O}M_{H_{2}O}}\; .
\end{equation} \\
The  division by 2 in this equation accounts for a pair of ions 
and $M_{H_{2}O}$ =18 g mol$^{-1}$. The Figure~\ref{molal}
 gives the value of the molality for each point of calculus
 \begin{table}[h]
\caption{Composition of NaCl solutions used in the simulations  
at 298.15 K and $1\;bar$. }
\label{molal}
\begin{center}
\begin{tabular}{|ccc|}
\hline\hline
Molality (m)  & $N_{H_{2}O}$ & $N_{ions}$ \\
\hline
0.06&862&2\\
0.99&832&32\\
1.99&806&58\\
3.07&778&86\\
4.05&754&110\\
5.0&732&132\\
5.93&712&144\\
6.02&710&154\\
6.31&704&160\\
	\hline
\end{tabular}
\end{center}
\end{table}

The static dielectric constant is computed from the 
fluctuations~\cite{neumann} of the total dipole moment {\bf M},
\begin{equation}
\epsilon=1+\frac{4\pi}{3k_BTV} (<{\bf M}^2>-<{\bf M}>^2)
\end{equation}
\noindent where  $k_B$ is the Boltzmann constant and $T$ the absolute 
temperature. The dielectric constant is obtained for long simulations at 
constant density and temperature or at constant temperature and pressure.
The shear viscosity is obtained using the autocorrelation function of the 
off-diagonal components of the pressure tensor $P_{\alpha\beta}$ according 
to the Green-Kubo formulation,
\begin{equation}
\eta= \frac{V}{k_BT} \int_0^\infty <P_{\alpha\beta}(t_0)P_{\alpha\beta}
(t_0+t)>_{t_0}dt,
\end{equation}\\
The self-diffusion coefficient, $D$ is obtained from the Einstein equation
\begin{equation}
D= \lim_{t \rightarrow \infty} \frac{1}{6t}\left<|{\bf R}_i(t)-
{\bf R}_i(0)|^2\right>,
\end{equation}
\noindent where ${\bf R}_i(t)$ is the center of mass position of 
molecule $i$ at time $t$ and $< ... >$ denotes time average.

Even thought the 
thermodynamic and dynamic
quantities were produced for  $864$ number of
particles for each density,  systems with $1024$ and $ 2048$
were also tested showing a difference in the result
smaller than the data points used in the plots. For
the solubility computations, where
the errors are larger,  $ 2048$ particles were
employed. In this particular case the error bars
are of the order of $2.7\%$ of the computed values.

\section{Results}

\subsection{ The Pure Sodium chloride NaCl/$\epsilon$}

\begin{figure}
\psfig{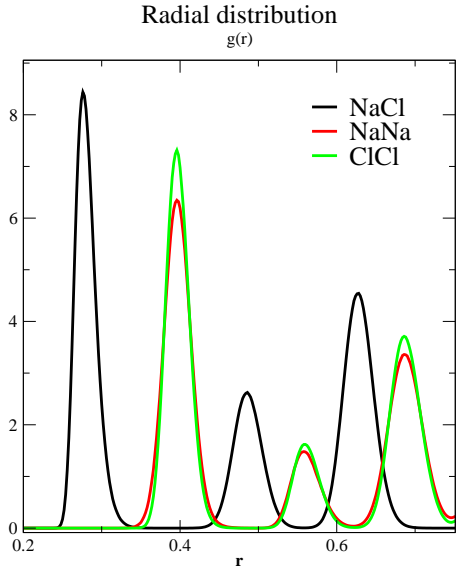}
\caption{ Radial distribution function g(r) versus the distance r 
at $1$bar and $298\;K$ for: Na-Na (red
line), Cl-Cl (green line) and Na-Cl (black line).  }
\label{Figa}
\end{figure}

The pure NaCl is analyzed. 
First, the parameters for the 
model were fitted to give the experimental value
for the  density of the NaCl crystal 
at the temperature of 
$T=298\;k$  and at the pressure of $1\;atm$ 
namely  $2.16 \;g \;cm^{-3}$~\cite{CRC}. Within
the parameters values which produce
this density, we select the subset
that also gives the  radial distribution for
Na-Na, Cl-Cl and Na-Cl as  illustrated
in the Figure~\ref{Figa}. This result
 shows  a peak in the curve
for the Na-Cl at 
$2.78\AA$ in agreement with the experiments~\cite{CRC}. 

Following this
procedure the resulting
 lattice energy (LE)  for the  NaCl/$\epsilon$ model
 is $669.21\;kJ/mol$ while 
the experimental data gives $790\;kJ/mol$~\cite{CRC}. The lattice
constant (LC) for the same model is $5.56\;\AA$ while the 
experimental value is $5.64\;\AA$~\cite{CRC}.

The reason for the difference between the
values for the lattice energy obtained within our approach
and the experiments is due to the ``screening'' factor
$\lambda$. In order to be consistent
with the idea that the $\lambda$ works 
as a screening the lattice energy
for the  NaCl/$\epsilon$ model
should  be computed in a renormalized form what it
will be explored in a future publication.

In order to validate our model, the Table~\ref{FFcomp} shows 
the values for the density, the lattice energy and the lattice constant
for the NaCl/$\epsilon$ in comparison with other force fields all
at $1\;bar$ of pressure and $298\;K$ of temperature.
While the  Alejandre et al.~\cite{alj09}(ACB) gives
 good results for the density, the JJ~\cite{JJ} model and 
the two parametrizations of the JC~\cite{JC08} approach(JC$_{S3}$ 
 and JC$_{T4}$)  show
good results for the lattice crystal and for the lattice energy when compared
with the experiments. Our model gives
good agreement with the experiments~\cite{CRC} for the density of the 
crystal and for the lattice constant, but
is a quite far from the reproduction of the lattice energy 
probably due to the way the lattice energy should
be modified for the NaCl/$\epsilon$ as explained above.

\begin{table}[h]
\caption{Density of NaCl at room pressure and temperature,
Lattice Energy, Lattice  Constant 
of various force fields and for experiments~\cite{CRC}.}
\label{FFcomp}
\begin{center}
\begin{tabular}{|cccc|}
\hline
 Model Ions& $\rho/(g/cm^3)$ & LC/\AA & LE/(kcal/mol) \\
\hline
JJ~\cite{JJ}& 1.78 &5.9&796.26\\
JC$_{S3}$~\cite{JC08} &1.97&5.7 &800.4\\
JC$_{T4}$~\cite{JC08}&2.05&5.78& 792.88\\
ACB.~\cite{alj09}& 2.16&5.47&816.37\\
this work & 2.16& 5.56&669.21\\
experimental~\cite{CRC}  &2.16&5.64&789.95\\
\hline
\end{tabular}
\end{center}
\end{table}

\begin{figure}
\centerline{\psfig{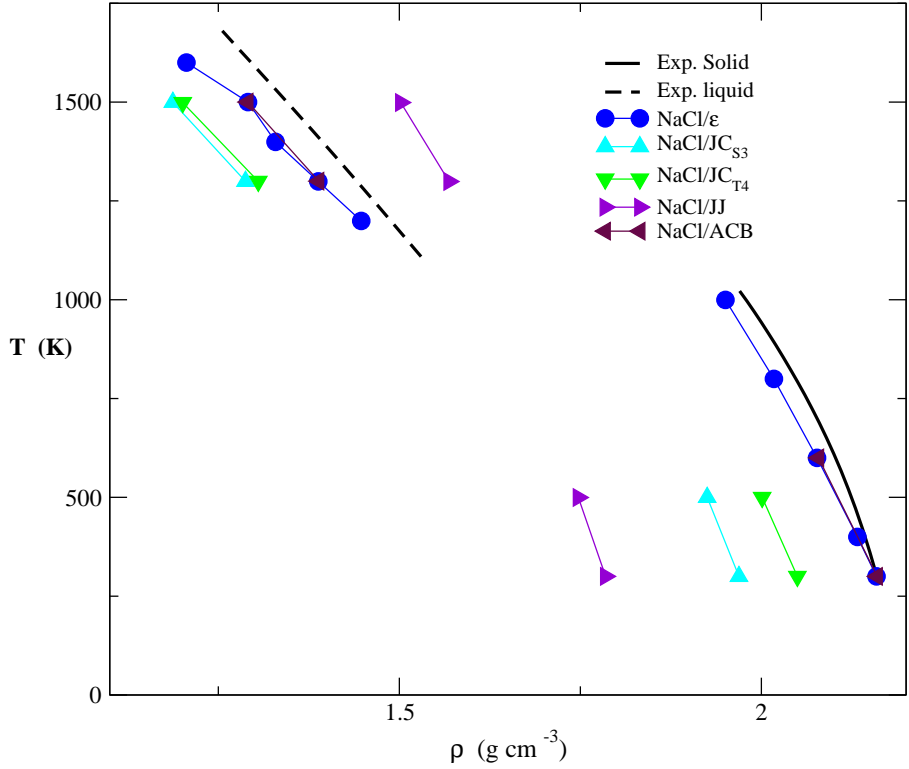}}
\caption{Temperature versus density of NaCl
  at the liquid 
(T>1000 K) and solid (T < 1100 K) phases. The solid and dashed black 
lines are the experimental data~\cite{CRC}, the blue
 filled circles are for the NaCl/$\epsilon$ model.
The results for the  ACB~\cite{alj09} model are represented by  brown filled 
triangles, for the 
JJ~\cite{JJ} model are shown by purple filled triangles, for 
the JC$_{S3}$~\cite{JC08} model
by blue triangles and for the JC$_{T4}$ model are shown as green triangles.}
\label{Fig2}
\end{figure}

Another important validation for the NaCl/$\epsilon$ is to check if the
 density of
the liquid phase, for temperatures higher than
the region from which the fitting was done, agrees
with the experimental results.
Figure~\ref{Fig2} illustrates the isobar 
at $1\;bar$ for the density versus temperature for
the system both in the solid and liquid phases. Our
results for the pure
 NaCl/$\epsilon$ (solid circles) are compared with 
the experimental data
(solid and dashed lines)~\cite{CRC} and with
 Alejandre et al.~\cite{alj09}(ACB) the JJ~\cite{JJ} model and 
the two parametrizations of the JC~\cite{JC08}(JC$_{S3}$  and JC$_{T4}$).
The NaCl/$\epsilon$ shows a better agreement with
the experiments than the other simulations.

\begin{figure}
\centerline{\psfig{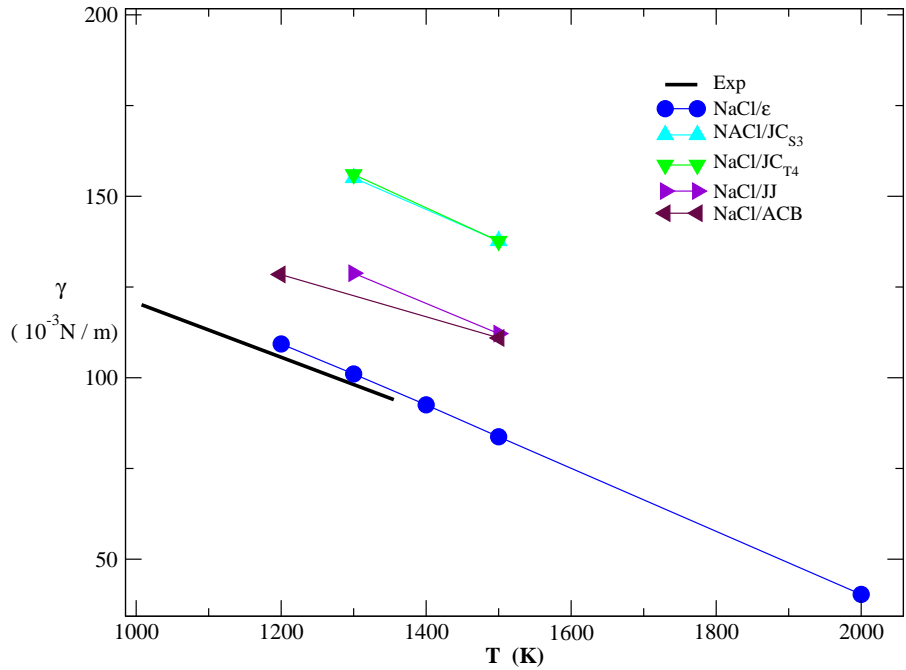}}
\caption{Surface Tension versus temperature for
the pure NaCl system at $1\;bar$ of pressure. The
 black line is the experimental 
data~\cite{CRC} and the blue circles for the NaCl/$\epsilon$.
The results for the  ACB~\cite{alj09} model are represented by  brown filled 
triangles, for the 
JJ~\cite{JJ} model are shown by purple filled triangles, for 
the JC$_{S3}$~\cite{JC08} model
by blue triangles and for the JC$_{T4}$ model are shown as green triangles.}
\label{Fig3}
\end{figure}

In addition to the density, the surface tension  was
also computed. Figure~\ref{Fig3} illustrates the temperature
versus surface tension for the NaCl/$\epsilon$  compared with  the experiments 
and the other models.  Our
model shows a better agreement with the
experimental surface tension when
compared with the  ACB~\cite{alj09} 
JJ~\cite{JJ}  JC$_{S3}$~\cite{JC08} and  JC$_{T4}$~\cite{JC08} models.

\subsection{ Sodium chloride NaCl/$\epsilon$ 
in the TIP4P/$\epsilon$ water}

\begin{figure}
\centerline{\psfig{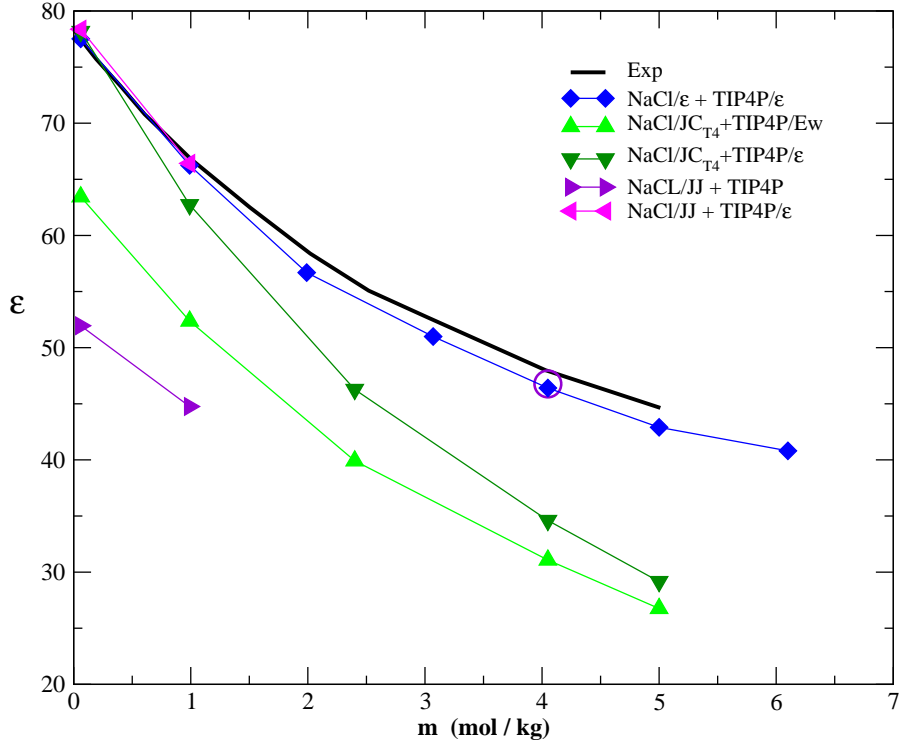}}
\caption{Dielectric constant of the 
mixture versus molal
concentration of the salt 
at the temperature  $298\;K$ and at $1\;bar$ of pressure. The 
black line is the experimental 
data~\cite{CRC}, the blue filled diamond 
is the results for the NaCl/$\epsilon$, the violet circle is 
the concentration in which 
the parametrization of our model was made. 
The purple and the dark
blue triangles are for the JJ salt model in the  TIP4P/$\epsilon$ 
and in the TIP4P  water 
models respectively.
The light and dark
green triangles are  for the  JC$_{T4}$ salt model
in the TIP4P$/Ew$
and in the TIP4P/$\epsilon$ water models respectively.All
the simulations have been performed in this work.}
\label{Fig4}
\end{figure}

The  thermodynamic and dynamic properties
of the NaCl/$\epsilon$  in solution with 
the  TIP4P/$\epsilon$ water
are checked
against experiments and other models.
The figure~\ref{Fig4} illustrates the
dielectric constant versus salt molal concentrations
at $298\;K$ and $1$bar
for the  NaCl/$\epsilon$ in the
TIP4P/$\epsilon$ water model (blue diamond) compared 
with the experimental data (solid black line)~\cite{CRC},
with the JJ~\cite{JJ}
 salt model in the  TIP4P/$\epsilon$  water (purple triangles) 
and in the TIP4P  (dark blue
triangles) water models respectively. In 
the case of the  JJ~\cite{JJ}
 salt model the simulations show 
phase separation for salt concentrations 
above $1\;molar$, so the dielectric
constant was not computed. The figure also
presents results
for  the  JC$_{T4}$~\cite{JC08} model
in the TIP4P$/Ew$ water  (dark green triangles)
and in the TIP4P/$\epsilon$ water respectively (light green triangles).
Our results indicates that even thought the parameters
for  the  NaCl/$\epsilon$  model were fitted to give the experimental
dielectric constant for 
the concentration $4$mol/kg (shown in the figure
as a purple circle), the  model 
in the TIP4P/$\epsilon$ water gives good  agreement with the experiments
over a wider range of molal concentrations.

\begin{figure}
\centerline{\psfig{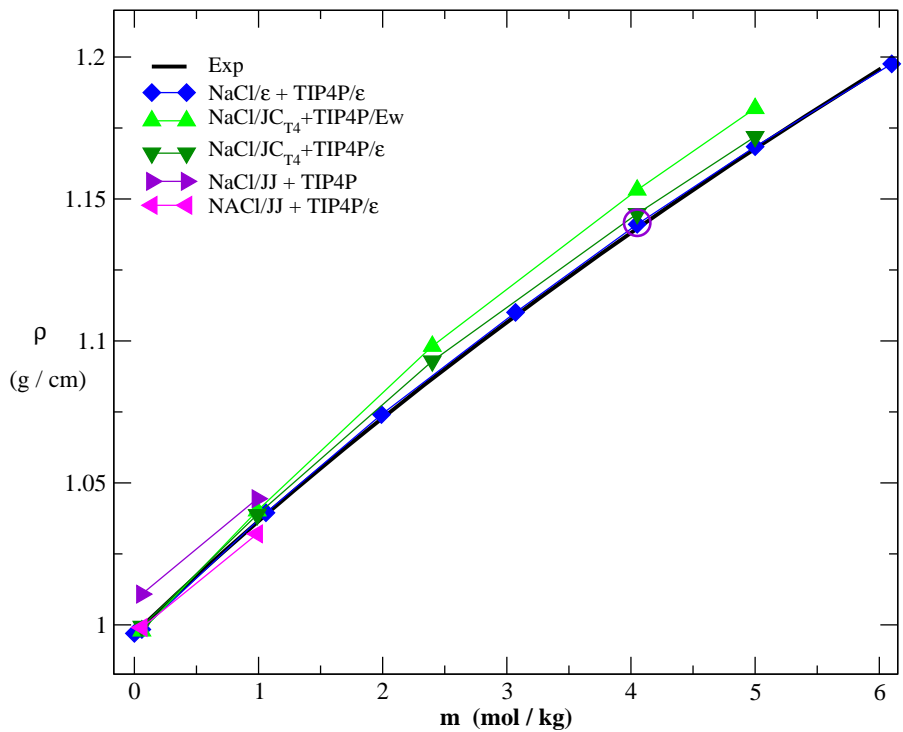}}
\caption{ Density of the mixture
versus molal concentration of the 
salt at the temperature of $298\;K$ and at $1$bar
and pressure. The black line is the experimental data~\cite{CRC} 
and the blue filled diamonds are the results for the
NaCl/$\epsilon$  model, the
violet circle
 is the diluted concentration where the parametrization was made. The purple 
and the dark blue triangles are  for the JJ model in 
the  TIP4P/$\epsilon$ 
and in the TIP4P  water models respectively.
The light and dark green triangles are the results for the  JC$_{T4}$ 
salt model
in the TIP4P$/Ew$
and in the TIP4P/$\epsilon$ water models
 respectively.}
\label{Fig5}
\end{figure}

Next, the Figure~\ref{Fig5} shows
the  density of the mixture of NaCl and water 
as a function of the salt molal concentration at 
$1\;bar$ and $298\;K$ for our model, other models
and the experimental results. Since most models are 
parametrized to give the correct experimental
density, for all the models presented in this
paper the agreement with the experiments are 
good.

\begin{figure}
\psfig{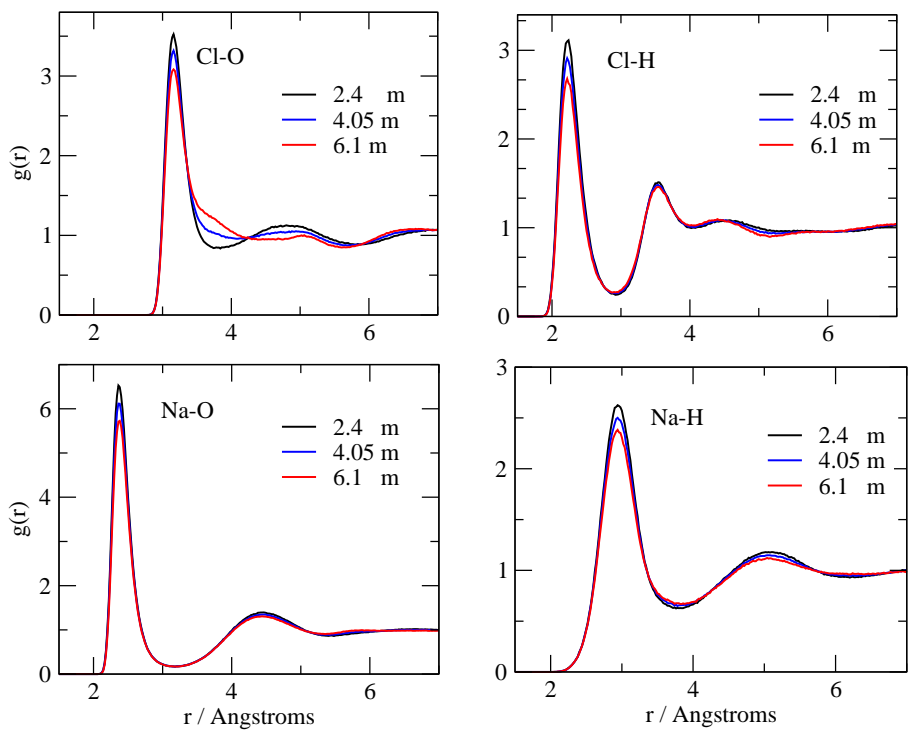}
\caption{Ion-water pair distribution functions using the rigid
 water model TIP4P/$\epsilon$  and NaCl/$\epsilon$ force field at 
$298\;K$, $1$ bar, and ionic concentrations of 
$2.4$ (black line), $4.05$ (blue line) and $6.1$ (red line) molal 
in all cases 864 molecules were used. }
\label{gr}
\end{figure}

Since our model introduces 
a new parameter related to the 
hydration of the ions, the  structure of the  water molecules
 around ions is computed and 
checked with experimental results.
The hydration is measured
 by the four partial pair distribution functions g$_{NaH}$,
g$_{ClH}$, g$_{NaO}$, and g$_{ClO}$ \cite{alj09}. The calculus of 
these functions with the
TIP4P/$\epsilon$  and NaCl/$\epsilon$ force fields are 
shown in  the Fig.~\ref{gr}   for a molal concentrations of 
$2.4, 4.05$ and $6.1$ molal. We can see that as the salt is diluted 
in water, the structure is favored around each ion, as
illustrated by the larger amplitudes of the first peaks, which
 points to a slightly stronger hydration. Having a smaller number of 
coordinating the anion has more mobility, as seen in the calculation 
of the diffusion coefficient.

The peak positions, $r_{max}$, 
of the pair distribution functions in our model are given by:
$r_{max}$ $\cong$ 3.0 \AA \ for Na-H,
$r_{max}$ $\cong$ 2.25 \AA \ and 3.55 \AA \ for the first and second peaks of
 Cl-H, $r_{max}$ $\cong$2.37 \AA \ for Na-O and 3.19 \AA \ for Cl-O. These
 values are in good agreement with experimental
 data~\cite{Jong,Powell,Skipper} namely, $r_{max}$$\cong$2.3 \AA \ 
first peak and $r_{max}$$\cong$3.7 \AA \ second peak for Cl-H; 
$r_{max}$$\cong$2.4 \AA \ for Na-O and $r_{max}$$\cong$3.2 \AA \ for Cl-O.

The water coordination numbers around the Na and Cl
 ions can be estimated by integrating the area under the first
 peak of the Na-O and Cl-O pair distribution functions up to
 the first minimum respectively. These coordination numbers 
are shown in the table~\ref{ncoord} and give a good
agreement with the experiments.

\begin{table}[h]
\caption{Ion-Water Coordination Numbers 
obtained by our simulations and experiments.  The uncertainties of 
experimental data~\cite{Mancinelli} are reported within parenthesis, along 
with the r-range used in the integration.}
\label{ncoord}
\begin{center}
\begin{tabular}{|c|cc|cc|}
\hline
molal& MD & MD &Exp~\cite{Mancinelli} &Exp~\cite{Mancinelli} \\
 concentration	&	NaO	&	ClO&NaO	&	ClO	 \\
\hline
\hline
2.4	&	4.75	&	6.55	& 4.83	(0.9)&6.68	(1.1) \\
4.05	&	4.55	&	6.4	 & 4.55	(1.4)&6.5	(1.3)\\
6.1	&	4	&	6.15	&-&- \\
\hline
\end{tabular}
\end{center}
\end{table}

\begin{figure}
\centerline{\psfig{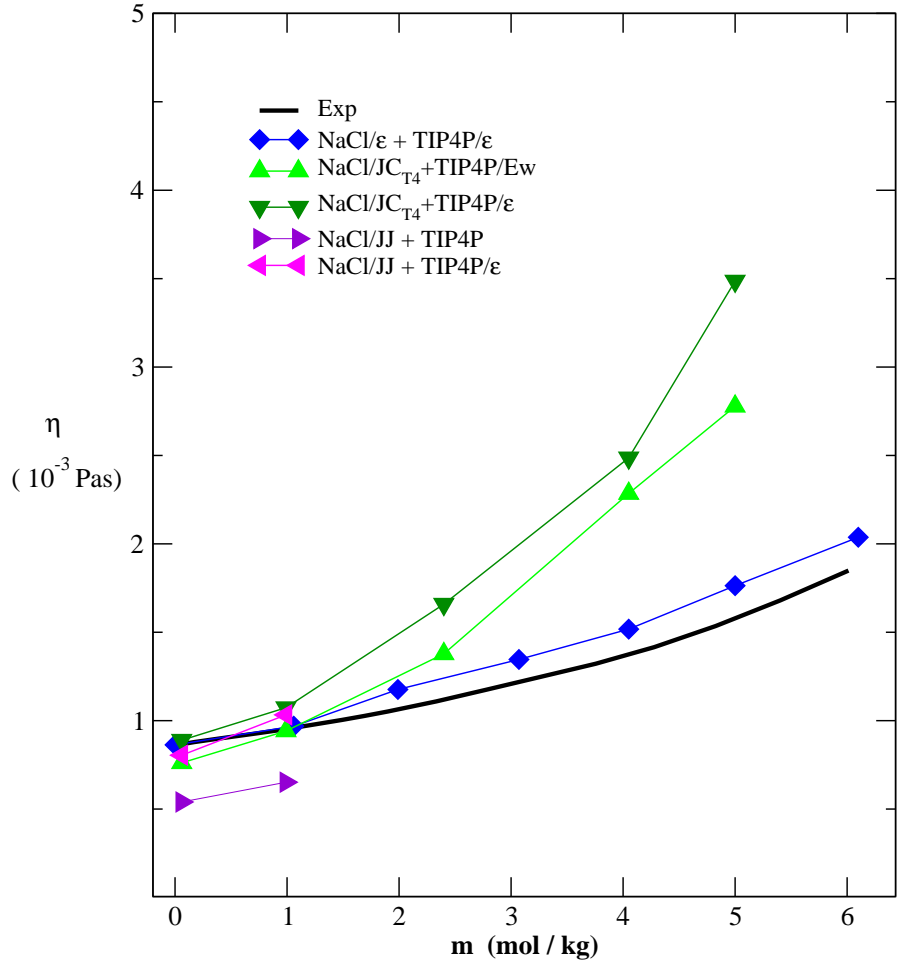}}
\caption{Viscosity  of the  NaCl molecules
immersed in water versus molal
concentration of the salt
at the temperature of $298\;K$ and at  $1\;bar$ of pressure.The black
 line is the experimental data~\cite{CRC} 
and the blue filled diamonds are the results for the
NaCl/$\epsilon$  model. The purple 
and the dark blue triangles are
 for the JJ model in the  TIP4P/$\epsilon$ 
and in the TIP4P water models 
respectively.
The light and dark green triangles are the results for 
the  JC$_{T4}$ model
in the TIP4P$/Ew$ 
and in the TIP4P/$\epsilon$ water models respectively. All the simulations
have been performed in this work.}
\label{Fig6}
\end{figure}

In addition to the thermodynamic 
functions already tested, it
is important to validate our 
model with dynamic properties. 
Then, the
 shear viscosity,  $\eta$,  of the  NaCl molecules
immersed in water at different 
molal concentrations, at $289\;K$ and at $1\;bar$  was 
evaluated. Figure~\ref{Fig6} illustrates the viscosity
versus molal concentration of the salt showing 
an increase of $\eta$ as the salt concentration
increases. This suggests
that the addition of 
salt makes the system 
more viscous. Our result is consistent with
the experimental values~\cite{CRC} and
show better agreement with the
experiments when compared 
with the JJ~\cite{JJ} model in the  TIP4P/$\epsilon$  
and in the TIP4P   water models
 respectively. The figure also compares 
our findings with the  JC$_{T4}$~\cite{JC08} model
in the TIP4P$/Ew$ water (light
green
triangles)and in the TIP4P/$\epsilon$ water respectively  (dark
green
triangles) indicating that the  NaCl/$\epsilon$  shows
a better performance.

Another important aspect of the dynamics of 
the particles is the diffusion. In this 
particular case it is interesting to 
observe how the water and the two 
ions change their mobilities with
the increase of the salt concentration.
This analysis can provide a good picture 
of the hydration process. 

\begin{figure}
\centerline{\psfig{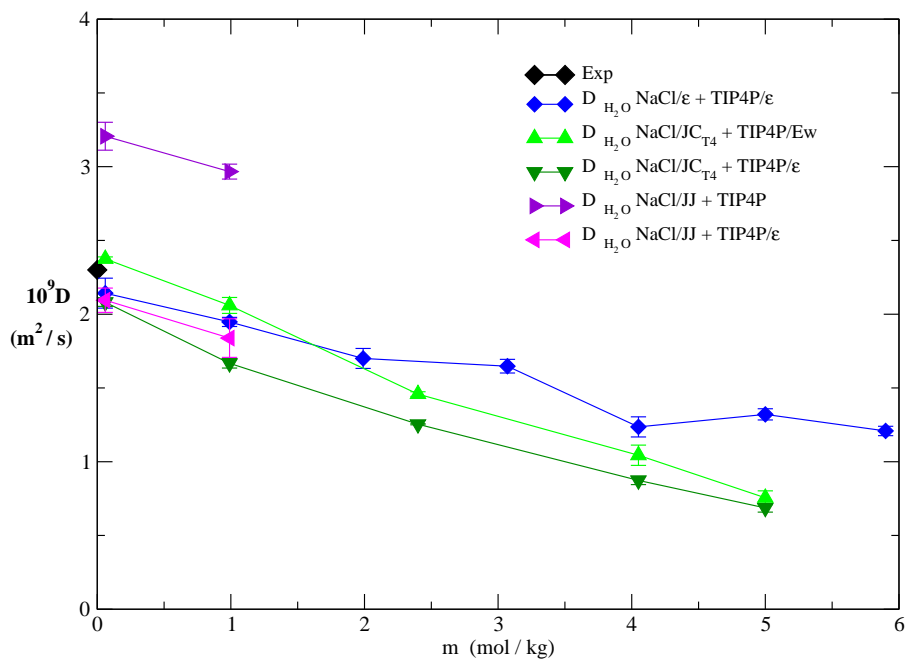}}
\caption{Diffusion coefficient of the water versus
molal concentration  of the salt at the temperature of $298\;K$ 
and at $1\;bar$ of 
pressure. The filled black diamond
 is the experimental data~\cite{Kuma} 
and the blue filled diamonds are the results for our model. The purple 
and the dark blue triangles are
 for the JJ model in the  TIP4P/$\epsilon$ 
and in the TIP4P water models 
respectively.
The light and dark green triangles are the results for 
the  JC$_{T4}$ model
in the TIP4P$/Ew$ 
and in the TIP4P/$\epsilon$ water models respectively. All
the simulations have been performed in this work}
\label{Fig7}
\end{figure}

In the Figure~\ref{Fig7} the self diffusion 
coefficient of water was measured for various salt 
concentrations at the temperature of $298\;K$ and at  $1\;bar$ of pressure. 
The filled black diamond in this figure
 shows the experimental data~\cite{CRC} 
and the blue filled diamonds are the results for our model. The purple 
and the dark blue triangles in the Figure~\ref{Fig7} are
 the results
for the JJ~\cite{JJ} model in the  TIP4P/$\epsilon$ 
and in the TIP4P water models 
respectively.
The light and dark green triangles in the same figure are the results for 
the  JC$_{T4}$~\cite{JC08} model
in the TIP4P$/Ew$  
and in the TIP4P/$\epsilon$ water models respectively.
As the molal concentration of 
the salt  increases the mobility
 of water molecules decreases. This behavior is consistent
with the idea that as the concentration of salt increases,
there are less particles of 
water free. The water
molecules are  hydrating the ions,
what slows down the dynamics of water. This result is 
in agreement with the increase of the
viscosity illustrated in the Figure~\ref{Fig6}.
\\

\begin{figure}
\centerline{\psfig{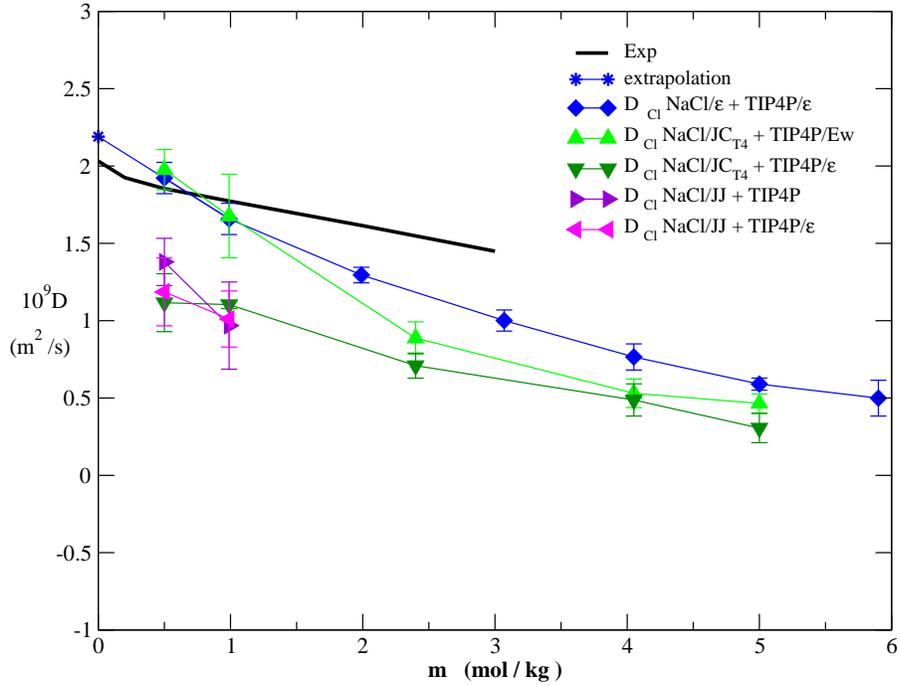}}
\caption{ Diffusion coefficient of chloride versus molal concentration
of the salt at the temperature of $298\;K$ and at $1\;bar$ of pressure. The 
 black line is the experimental data~\cite{Kuma} and the blue
 filled diamonds are the results for the NaCl/$\epsilon$
 model.The purple 
and the dark blue triangles are
 for the JJ model in the  TIP4P/$\epsilon$ 
and in the TIP4P water models 
respectively.
The light and dark green triangles are the results for 
the  JC$_{T4}$ model
in the TIP4P$/Ew$
and in the TIP4P/$\epsilon$ water models respectively. All
the simulations have been performed in this work.}
\label{Fig8}
\end{figure}

The diffusion coefficient of the chloride ions versus
molal concentration  of the salt at the temperature of $298\;K$ 
and at $1\;bar$ of 
pressure is shown in the 
Figure~\ref{Fig8}. The black line in this figure
indicates the experimental data~\cite{CRC} 
and the blue filled diamonds are the results for our model. The purple 
and the dark blue triangles in the Figure~\ref{Fig8}  are
the results
 for the JJ~\cite{JJ} model in the  TIP4P/$\epsilon$ 
and in the TIP4P  water models 
respectively.
The light and dark green triangles in the 
same figure  are the results for 
the  JC$_{T4}$~\cite{JC08} model
in the TIP4P$/Ew$ 
and in the TIP4P/$\epsilon$ water models respectively. The 
experimental data  at 
infinite dilution of diffusion coefficient 
is $D_{Cl}=$ 2.032 $10^{-5} cm ^2 s^{-1}$. 
The system shows a decrease in mobility with
the increase of the concentration what is 
the natural behavior of any molecular system. The 
increase of the number of particles, decreases the mobility.

\begin{figure}
\centerline{\psfig{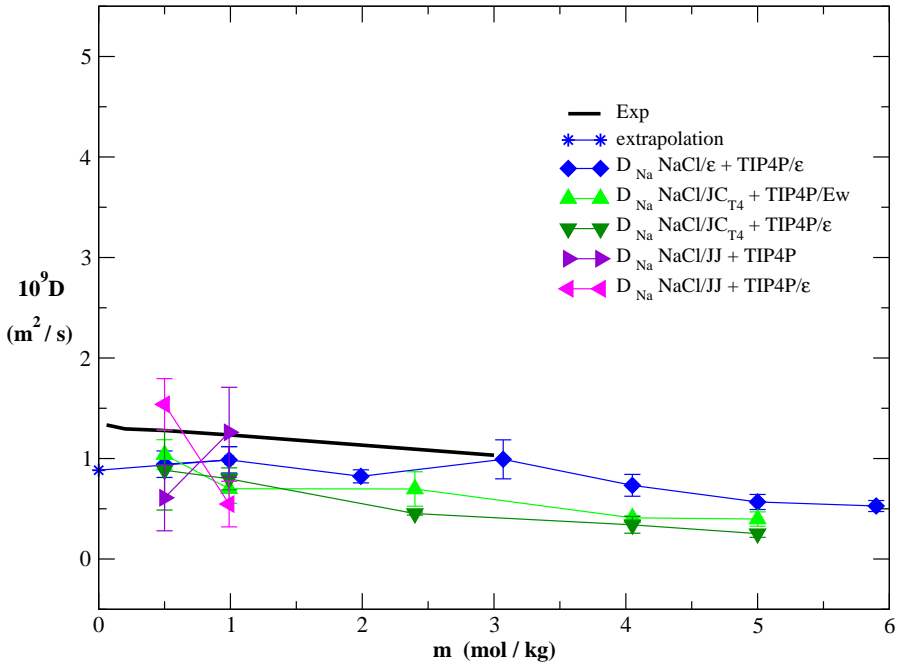}}
\caption{Diffusion coefficient
of Na versus the molal concentration of the salt
at temperature of $298\;K$ and at $1\;bar$ of pressure. The 
 black line is the experimental data~\cite{Kuma} and the 
blue filled diamonds are the results for
the NaCl/$\epsilon$ model.The purple 
and the dark blue triangles are
 for the JJ model in the  TIP4P/$\epsilon$ 
and in the TIP4P water models 
respectively.
The light and dark green triangles are the results for 
the  JC$_{T4}$ model
in the TIP4P$/Ew$
and in the TIP4P/$\epsilon$ water models respectively. All
the simulations have been performed in this work }
\label{Fig9}
\end{figure}

The diffusion coefficient 
of the sodium versus the molal concentration
of the salt is shown in the 
Figure~\ref{Fig9}. The black line in this figure illustrates
the experimental data~\cite{CRC} 
and the blue filled diamonds are the results for our model. The purple 
and the dark blue triangles in the same figure are the results
 for the JJ model in the  TIP4P/$\epsilon$ 
and in the TIP4P water models 
respectively.
The light and dark green triangles in the Figure~\ref{Fig9} 
are the results for 
the  JC$_{T4}$ model
in the TIP4P$/Ew$ 
and in the TIP4P/$\epsilon$ water models respectively. The diffusion
coefficient of the sodium  is almost constant when
the salt concentration is increased. This behavior
might be attributed to the small size of the hydrated sodium
when compared with the hydrated chloride.

\begin{figure}
\psfig{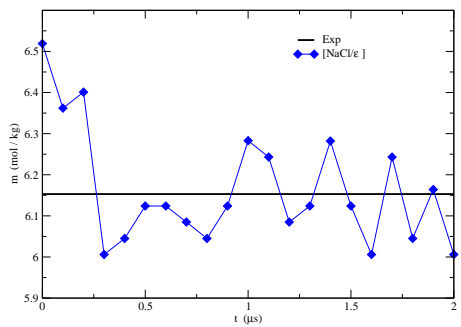}
\caption{   Solubility for the system  
 NaCl/$\varepsilon$  salt and Tip4p/$\varepsilon$ water 
at $1\; bar$ and $298\;K$  }
\label{fig:time}
\end{figure}

The solubility was computed
employing the method number four of the reference by 
Manzanilla-Granados  et al.~\cite{Hector} as follows. At the beginning of the simulation, a nano-crystal 
is dipped into a  6.5 mol kg$^{-1}$ solution of  
NaCl/$\varepsilon$  ions and Tip4p/$\varepsilon$ water. Values
for the solubility were computed 
for a time up to 2 microseconds, as shown in Figure~\ref{fig:time}
what indicates that the after $0.4\;\mu s$ the simulation
stabilizes with a fluctuation of $2.7\%$

\begin{figure}
\centerline{\psfig{figure=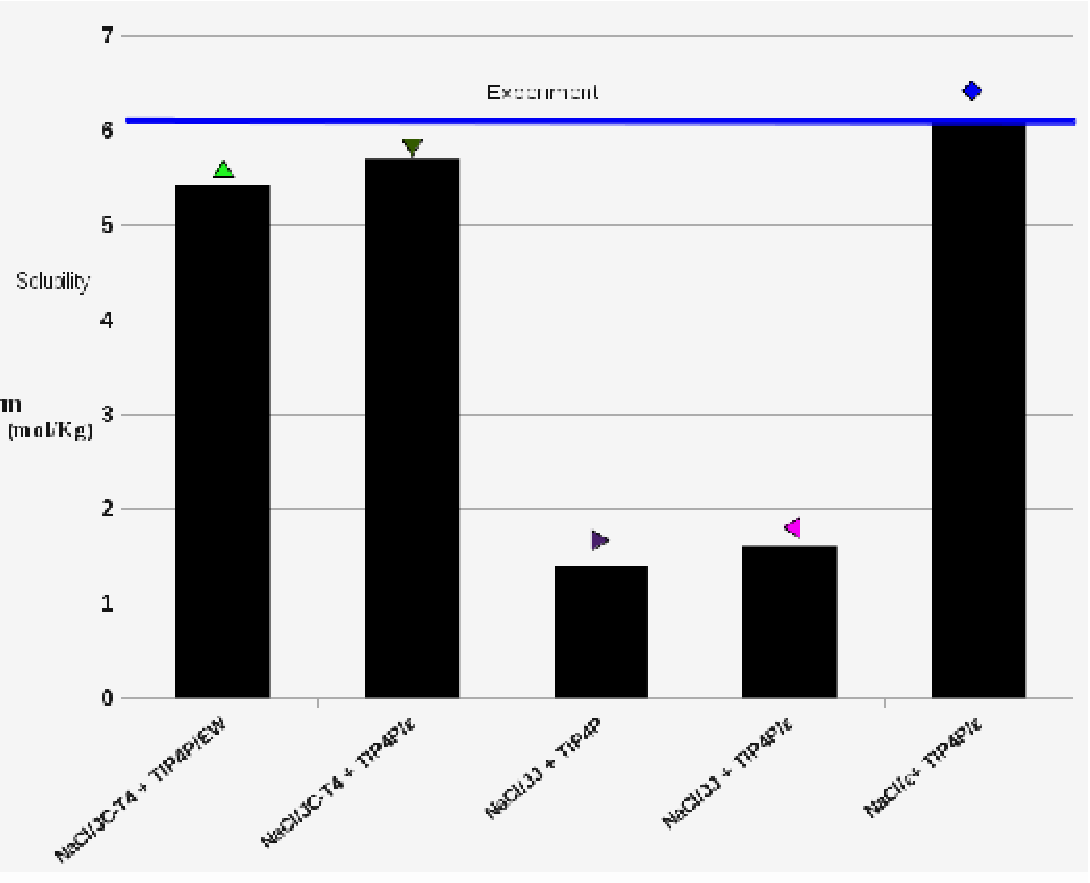,clip,width=12.0cm,angle=0}}
\caption{Solubility for $1\;bar$ of
pressure and $298\;K$ of temperature for:
the  NaCl/$\epsilon$ salt in the TIP4P/$\epsilon$ water,  for 
the JC$_{T4}$ salt 
in the TIP4P$/Ew$ water
and in the TIP4P/$\epsilon$   water, JJ salt
 in the  TIP4P/$\epsilon$ 
and in the TIP4P  water. Experimental results are
show as the blue line~\cite{CRC}. }
\label{Fig10}
\end{figure}

 The solubility for the 
NaCl/$\epsilon$ is compared in the Figure~\ref{Fig10} with 
the experimental results and with the results for other
models, showing that our
model has  better agreement with the experiments than the other models.

\subsection{NaCl/$\epsilon$ in the  SPC/$\epsilon$ water}

In order to further validate our model, we analyze the behavior of 
the NaCl/$\epsilon$ in a solution 
with a different water model. 
For this purpose the  SPC/$\epsilon$ was selected.
This  force field  reproduces very  the experimental
dielectric constant and
the density of pure water at various thermodynamic states. It fails,
however, to  reproduce the transport properties~\cite{spce}.

\begin{figure}
\centerline{\psfig{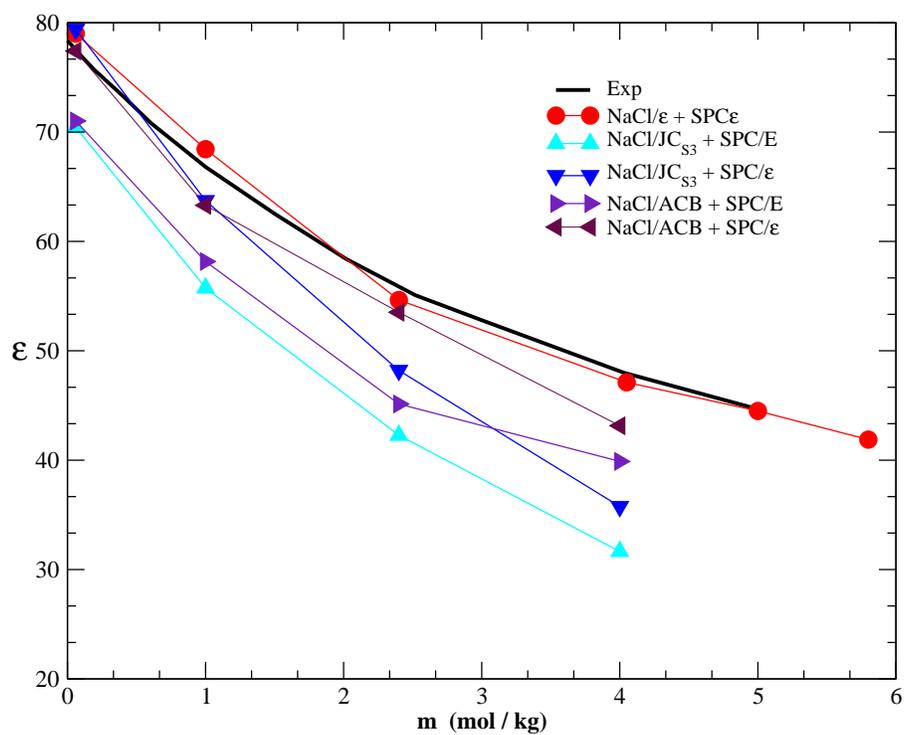}}
\caption{Dielectric constant versus molal concentration
of salt at temperature of $298\;K$ and $1\;bar$ of pressure
for the NaCl/$\epsilon$ model in  the   SPC/$\epsilon$ 
model for water(red circles), for
the experiments~\cite{CRC}
(solid black line), for the ACB~\cite{alj09} model
in the SPC/E water (dark blue triangles) and 
in the SPC/$\epsilon$ water (black triangles) and for the 
JC$_{S3}$~\cite{JC08} model 
in the  SPC/E water (light blue triangles) and 
in the SPC/$\epsilon$ water (blue triangles).}
\label{Fig11}
\end{figure}

First, Figure~\ref{Fig11} shows
the dielectric constant 
at $1\;bar$ of pressure
and at the temperature 
of $298\;K$ for different molal concentrations of salt
for the NaCl/$\epsilon$ model in  the   SPC/$\epsilon$ 
model for water(red circles), for
the experiments~\cite{CRC}
(solid black line), for the ACB~\cite{alj09} model
in the SPC/E water (dark blue triangles) and 
in the SPC/$\epsilon$ water (black triangles) and for the 
JC$_{S3}$~\cite{JC08} model 
in the  SPC/E water (light blue triangles) and 
in the SPC/$\epsilon$ water (blue triangles).
The graph shows that  $\epsilon$ decreases
as the concentration of salt increases  due
to the hydration effects as it
would be expected.  
.

\begin{figure}
\centerline{\psfig{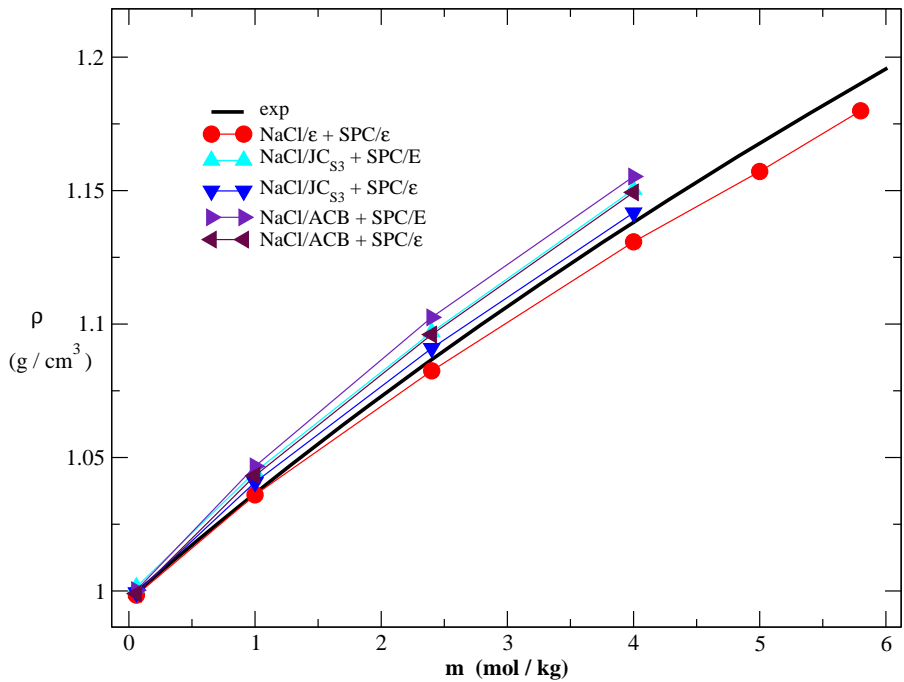}}
\caption{Density versus molal concentration of the salt at 
temperature of $298\;K$ and $1\;bar$ of pressure
for the NaCl/$\epsilon$ model (red circles), for
the experiments~\cite{CRC}
(solid black line), for the ACB~\cite{alj09} model
in SPC/E water (dark blue triangles) and 
in SPC/$\epsilon$ water (black triangles) and for the 
JC$_{S3}$~\cite{JC08} model 
in in SPC/E water (light blue triangles) and 
in SPC/$\epsilon$ water (blue triangles).} 
\label{Fig12}
\end{figure}

Next, the density was 
computed for different
molal concentrations of 
the salt.  Figure~\ref{Fig12} illustrates the density for 
the NaCl/$\epsilon$ model in  the   SPC/$\epsilon$ 
model for water(red circles), for
the experiments~\cite{CRC}
(solid black line), for the ACB~\cite{alj09} model
in the SPC/E water (dark blue triangles) and 
in the SPC/$\epsilon$ water (black triangles) and for the 
JC$_{S3}$~\cite{JC08} model 
in the  SPC/E water (light blue triangles) and 
in the SPC/$\epsilon$ water (blue triangles).
The NaCl/$\epsilon$ model in the SPC/$\epsilon$ 
underestimates the density. This
might be due to the fact that the SPC/$\epsilon$ water
model has a higher dipole moment when compared
with the TIP4P/$\epsilon$ model.\\

\begin{figure}
\centerline{\psfig{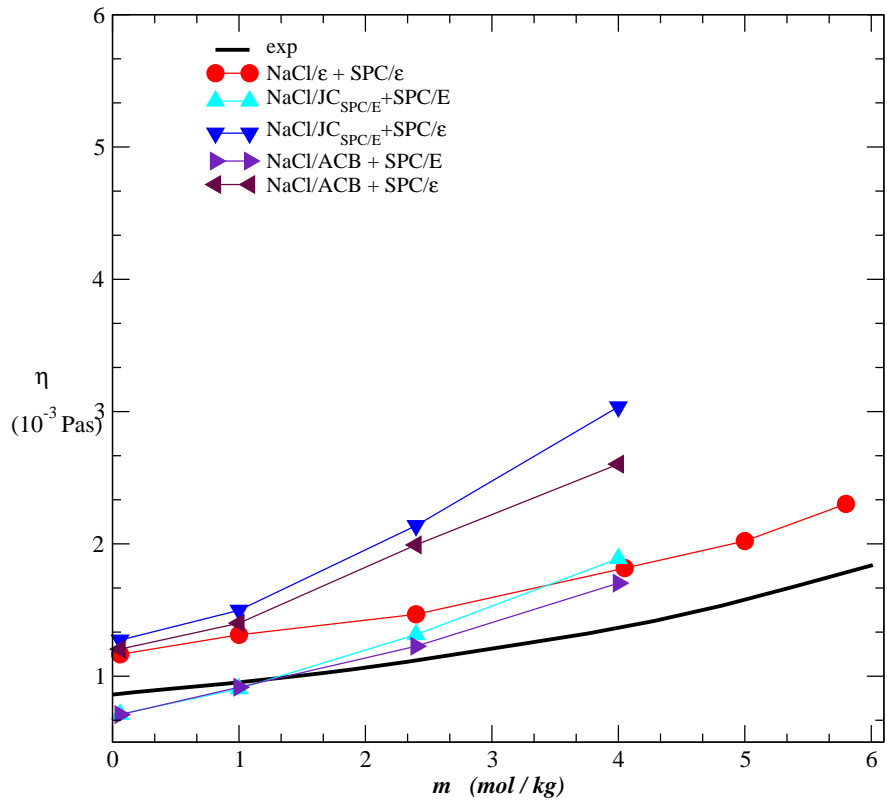}}
\caption{Shear viscosity versus molal concentration of salt 
at temperature of $298\;K$ and $1\;bar$ of pressure for 
the NaCl/$\epsilon$ model (red circles), for
the experiments~\cite{CRC}
(solid black line), for the ACB~\cite{alj09} model
in SPC/E water (dark blue triangles) and 
in SPC/$\epsilon$ water (black triangles) and for the 
JC$_{S3}$~\cite{JC08} model 
in in SPC/E water (light blue triangles) and 
in SPC/$\epsilon$ water (blue triangles).}
\label{Fig13}
\end{figure}

Then,  we also test the dynamics of the system.
Figure~\ref{Fig13} shows the viscosity,$\eta$,  versus
the molal salt concentration at $1\;bar$ of
pressure and $298\;K$ of temperature of
for the NaCl/$\epsilon$ model (red circles), for
the experiments~\cite{CRC}
(solid black line), for the ACB~\cite{alj09} model
in SPC/E water (dark blue triangles) and 
in SPC/$\epsilon$ water (black triangles) and for the 
JC$_{S3}$~\cite{JC08} model 
in in SPC/E water (light blue triangles) and 
in SPC/$\epsilon$ water (blue triangles).

The figure shows that the viscosity increases with the increase
of the salt concentration what can be attributed to the 
solvation. The 
values for our
model in the  SPC/$\epsilon$ water
 show  a shift in
the solubility   when compared with the experimental
results. The origin of this shift is probably related
to the fact that the SPC/$\epsilon$ does not 
perform well for dynamic properties. The constant
shift therefore might be due to the constant concentration
of water present in the solution.

\begin{figure}
\centerline{\psfig{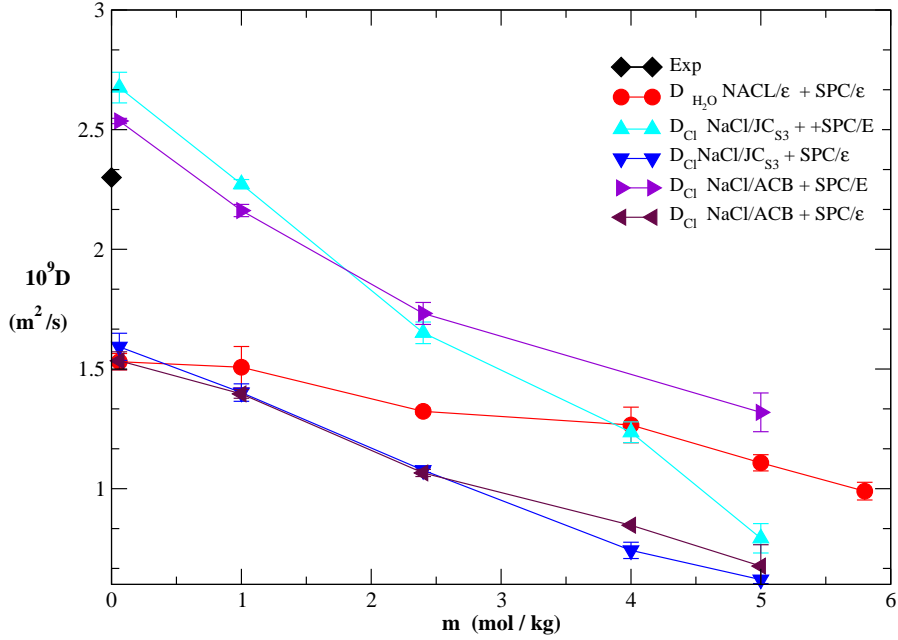}}
\caption{Diffusion coefficient of water versus
molal concentration of the salt at temperature of $298\;K$ and $1\;bar$ 
of pressure for the NaCl/$\epsilon$ model (red circles), for
the experiments~\cite{Kuma}
(black diamond), for the ACB~\cite{alj09} model
in SPC/E water (dark blue triangles) and 
in SPC/$\epsilon$ water (black triangles) and for the 
JC$_{S3}$~\cite{JC08} model 
in in SPC/E water (light blue triangles) and 
in SPC/$\epsilon$ water (blue triangles). }
\label{Fig14}
\end{figure}

In order to test if the incorrect dynamical behavior 
of the NaCl/$\epsilon$ and SPC/$\epsilon$ mixture
is due to the problems in the water model, the 
diffusion coefficient is also computed.
Figure~\ref{Fig14} illustrates the diffusion
coefficient of water versus the molal concentration
of the salt at room pressure and 
temperature. $D$ decreases 
with the increasing concentration of salt due
to the solvation effects. The mobility
for the NaCl/$\epsilon$ and SPC/$\epsilon$ mixture
for low concentrations of salt is
much lower than the diffusion coefficient
observed for the NaCl/$\epsilon$ and TIP4P/$\epsilon$
water model and far below the experimental results.

\begin{figure}
\centerline{\psfig{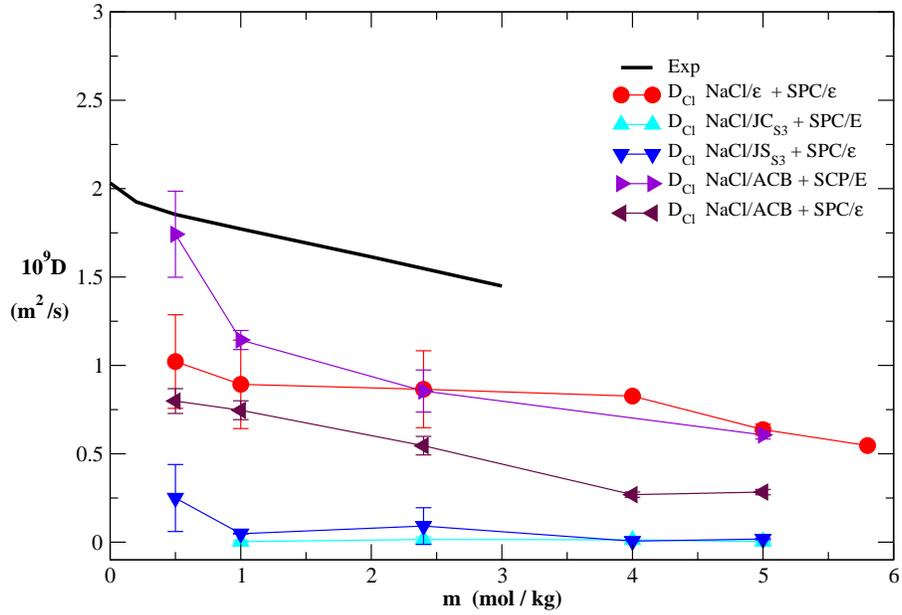}}
\caption{Diffusion coefficient of Cl versus
molal concentration of the salt at temperature of $298\;K$ 
and $1\;bar$ of pressure
for the NaCl/$\epsilon$ model (red circles), for
the experiments~\cite{Kuma}
(black solid line), for the ACB~\cite{alj09} model
in SPC/E water (dark blue triangles) and 
in SPC/$\epsilon$ water (black triangles) and for the 
JC$_{S3}$~\cite{JC08} model 
in in SPC/E water (light blue triangles) and 
in SPC/$\epsilon$ water (blue triangles).}
\label{Fig15}
\end{figure}

The  diffusion coefficient of the chloride is shown in  the 
Figure~\ref{Fig15} at temperature of $298\;K$ and $1\;bar$ 
of pressure for the NaCl/$\epsilon$ model (red circles), for
the experiments~\cite{Kuma}
(black diamond), for the ACB~\cite{alj09} model
in SPC/E water (dark blue triangles) and 
in SPC/$\epsilon$ water (black triangles) and for the 
JC$_{S3}$~\cite{JC08} model 
in in SPC/E water (light blue triangles) and 
in SPC/$\epsilon$ water (blue triangles). It shows an smooth decrease with 
the concentration of salt

\begin{figure}
\centerline{\psfig{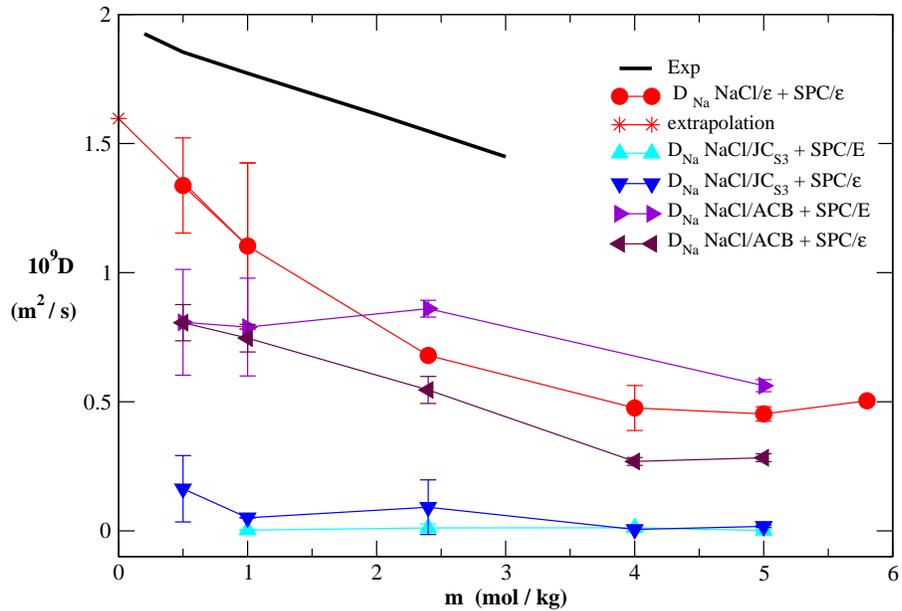}}
\caption{Diffusion coefficient of Na versus
molal concentration of the salt at temperature of $298\;K$ and $1\;bar$ of 
pressure
for the NaCl/$\epsilon$ model (red circles), for
the experiments~\cite{Kuma}
(black solid line), for the ACB~\cite{alj09} model
in SPC/E water (dark blue triangles) and 
in SPC/$\epsilon$ water (black triangles) and for the 
JC$_{S3}$~\cite{JC08} model 
in in SPC/E water (light blue triangles) and 
in SPC/$\epsilon$ water (blue triangles).}
\label{Fig16}
\end{figure}

Figure~\ref{Fig16} shows the diffusion 
coefficient for the sodium versus the salt concentration
at temperature of $298\;K$ and $1\;bar$ 
of pressure for the NaCl/$\epsilon$ model (red circles), for
the experiments~\cite{Kuma}
(black diamond), for the ACB~\cite{alj09} model
in SPC/E water (dark blue triangles) and 
in SPC/$\epsilon$ water (black triangles) and for the 
JC$_{S3}$~\cite{JC08} model 
in in SPC/E water (light blue triangles) and 
in SPC/$\epsilon$ water (blue triangles).

\begin{figure}
\centerline{\psfig{figure=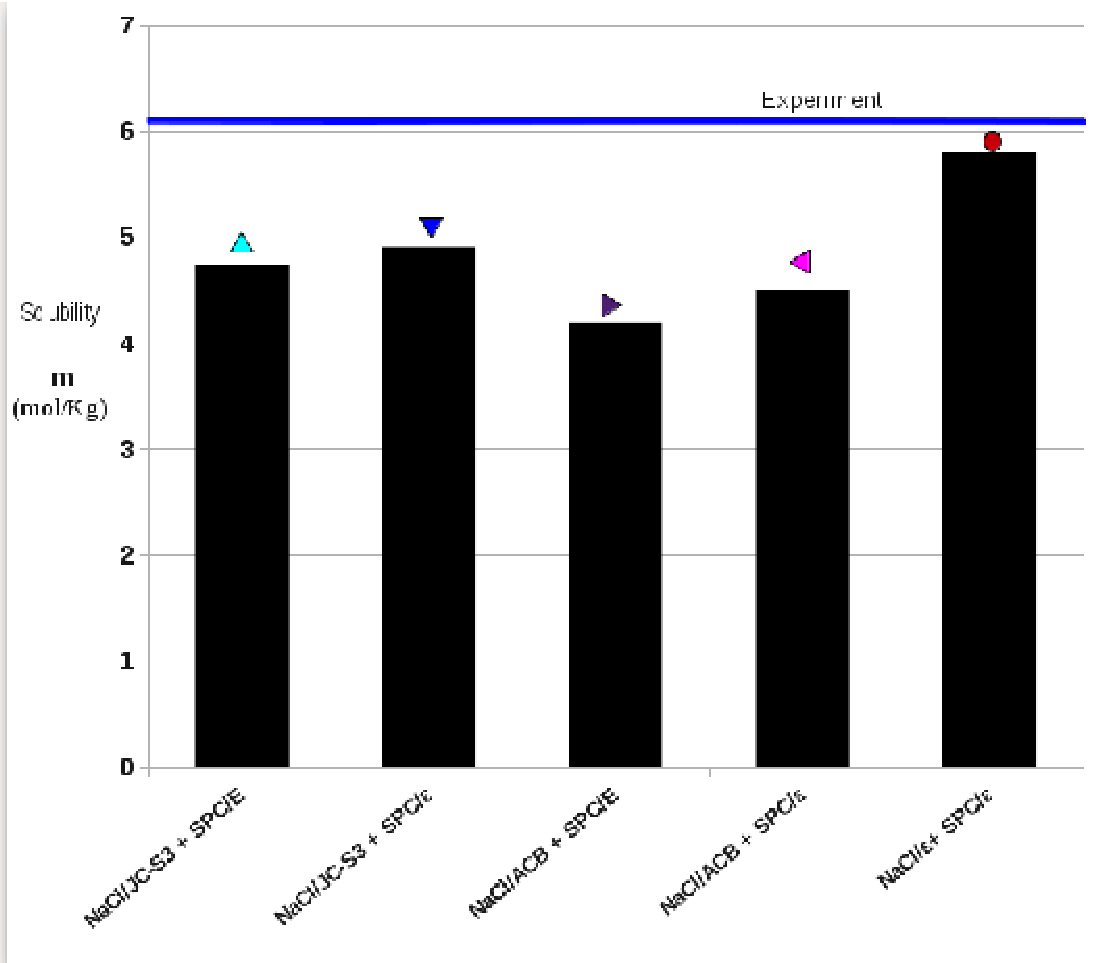,clip,width=12.0cm,angle=0}}
\caption{Calculus of the solubility using the direct coexistence method DCM, described by Manzanilla et al \cite{Hector}, at temperature of $298\;K$ and $1\;bar$ of pressure, for the NaCl/$\epsilon$ and SPC/$\epsilon$. The other values for SPC/$\epsilon$ were calculated in this work and the other are taken from the original work \cite{Jo09}.
The blue line is the experimental data~\cite{CRC}. }
\label{Fig17}
\end{figure}

Finally, the solubility 
was also computed. 
The  value obtained for the solubility 
for the NaCl/$\epsilon$ model in the SPC/$\epsilon$ water
is 5.8 mol kg$^{-1}$ with and error the $\pm$ 0.15 mol kg$^{-1}$ 
after 1$\mu$s of simulation. The other values for SPC/$\epsilon$ were calculated in this work and the other are taken from the original work \cite{Jo09}. The error bar is 
due to approximations in the method employed to calculate the 
solubility~\cite{Hector}. The  NaCl/$\epsilon$ model 
in the SPC/$\epsilon$ water is compared with experiments and with other 
salt models in the Figure~\ref{Fig17}, showing that
it underestimate the value of the solubility when 
compared with the result
for the solubility obtained with 
 the NaCl/$\epsilon$ model 
in the TIP4P/$\epsilon$ water.

\section{Conclusions}

In this paper, we have proposed  the NaCl/$\epsilon$
nonpolarizable model for NaCl. Within our 
approach the interaction potential of
 the ions combines a Lennard-Jones term
and a Coulombic potential. The combination
of the two terms is balanced by a parameter
$\lambda_i$ for each particle. 
The parametrization of our model
uses experimental results for both
the pure salt system and the 
mixture between the salt and water.
In this  process the 
water model selected for
the   salt-water mixture, the  TIP4P/$\epsilon$ 
water model,  shows the appropriated dielectric constant. 

Then  NaCl/$\epsilon$ model
 was validated by computing
the density, the dielectric constant, the 
surface tension, the diffusion and 
the viscosity for various concentrations
of the salt. Our results for
the pure salt system show a good
agreement with the experiments, particularly when
compared with the same quantities computed for other
salt models.

In addition, the mixture of the  NaCl/$\epsilon$ model
with the  TIP4P/$\epsilon$ water model was studied.
The density, dielectric constant, diffusion, solubility 
and viscosity were computed, showing a good
agreement with experiments when compared
with the results obtained using other salt and 
water models.

Finally, the NaCl/$\epsilon$ and  SPC/$\epsilon$ mixture
was analyzed. 
In this case, the thermodynamic quantities 
perform well, while the diffusion show discrepancies 
that in fact are consistent with the discrepancies 
of the bulk diffusion coefficient for this model.
Our results indicate that the combination of
the  NaCl/$\epsilon$ 
with the  TIP4P/$\epsilon$  models are
good  for describing
salt solutions.

\section {Acknowledgments}
 
 We thank the Brazilian agencies CNPq, INCT-FCx, and Capes for the
financial support. We also thank the CONACYT and SECITI of Mexico city for 
financial support.

\date{}

\begin{tocentry}
\includegraphics[width=7cm]{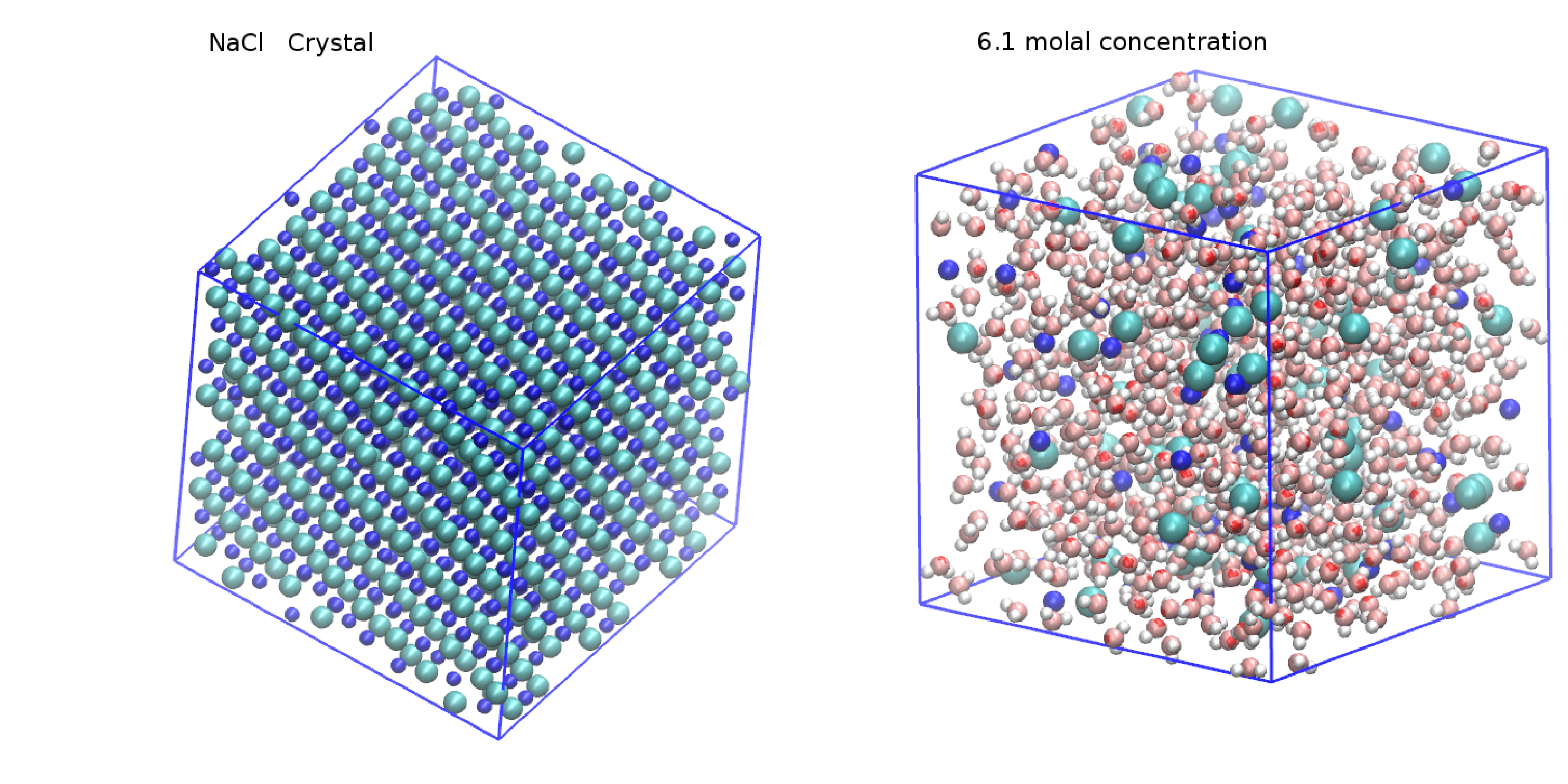}
\end{tocentry}

\end{document}